\documentclass{elsart}

\usepackage{amssymb}
\usepackage{graphicx}

\journal{Optical Communication}
\begin{document}

\begin{frontmatter}

\title{ The discrete fractional random cosine and sine transforms}

\author{Zhengjun Liu, Qing Guo and}
\author{Shutian Liu\corauthref{cor1}}
\ead{stliu@hit.edu.cn} \corauth[cor1]{Corresponding author}
\address{Harbin Institute of Technology, Department of Physics, Harbin 150001 P. R. CHINA}

\begin{abstract}
Based on the discrete fractional random transform (DFRNT), we
present the discrete fractional random cosine and sine transforms
(DFRNCT and DFRNST). We demonstrate that the DFRNCT and DFRNST can
be regarded as special kinds of DFRNT and thus their mathematical
properties are inherited from the DFRNT. Numeral results of DFRNCT
and DFRNST for one and two dimensional functions have been given.
\end{abstract}

\end{frontmatter}

\section{Introduction}

Recently we proposed a discrete fractional random transform
(DFRNT)\cite{Liu:2005} and have demonstrated its application to
image encryption and decryption. The DFRNT is a kind of discrete
transform with fractional order originated from the fractional
fourier transform (FrFT) \cite{ozaktas:2000} and especially the
discrete fractional Fourier transform (DFrFT)\cite{pei:1997}, and
thus has the most excellent mathematical properties as FrFT and
DFrFT have. Meanwhile, however, the result of transform itself can
be inherently random.

Because the close relationship of the DFRNT and the DFrFT(FrFT),
we hope that the DFRNT can serve as a general mathematical and
numerical tool in the field of digital signal processing. In the
first step towards this end, we extend the DFRNT to discrete
fractional random cosine and sine transforms (DFRNCT and DFRNST).
We know that the Cosine and Sine transforms and their discrete
versions are useful tools in signal and image processing, such as
signal coding\cite{bellifemine:1994},
watermarking\cite{Chang:2005} and restoration of de-focused
images\cite{Lam:1998}. In the Ref. \cite{pei:2001} Pei and Yeh
extended the Cosine transform to the discrete fractional cosine
transform (DFrCT) and the discrete fractional sine transform
(DFrST). Both of them possess well the angle additivity property
of the DFrFT. Moreover, the DFrCT and DFrST are used in the
digital computation of FrFT for the reducing computational load of
the DFrFT.

The motivation of this paper is to propose the definitions of the
DFRNCT and DFRNST directly based on the DFRNT and investigate
their mathematical properties. We find that the DFRNCT and DFRNST,
generated by the method of Pei and Yeh \cite{pei:2001}, are very
close to the original DFRNT. We can demonstrate that they are only
two subsets of DFRNT. From DFRNCT and DFRNST we can also
regenerate another DFRNT which will have nice symmetric properties
for even and odd signals. Numerical simulations have demonstrated
such properties.

In Section 2, we briefly introduce the definition of DFRNT and
give its mathematical properties. And then we give the definitions
of DFRNCT and DFRNST based on the DFRNT. Their mathematical
properties and the relationships with DFRNT are also given. In
Section 3 we give some numerical results. Conclusions are given in
Section 4.

\section{The DFRNT, DFRNCT and DFRNST}
\subsection{The DFRNT and its properties}

The DFRNT's of a 1-D and a 2-D signals, denoted by $x$ and $y$
respectively, can be written as matrix multiplications as
follows\cite{Liu:2005}
\begin{equation}
X_{\alpha}=\mathcal{R}^{\alpha}x,
\end{equation}
\begin{equation}
Y_{\alpha}=\mathcal{R}^{\alpha}y(\mathcal{R}^{\alpha})^t,
\end{equation}
where $\mathcal{R}^{\alpha}$ is the kernel transform matrix of the
DFRNT and can be expressed as
\begin{equation}
\mathcal{R}^{\alpha}=VD^{\alpha}V^t.
\end{equation}
In the kernel matrix, $D^{\alpha}$ is a diagonal matrix generated
by a set of values $\{\exp(-2i\pi n\alpha/M):n=0,1,2,...N-1\}$,
which are considered to be the eigenvalues of the DFRNT. Where
$\alpha$ indicates the fractional order of the DFRNT. $M$ is a
positive number, usually is an integer which has a meaning of
periodicity with respective to the fractional order $\alpha$ in
eigenvalues. $D^{\alpha}$ is written as follows
\begin{equation}
D^{\alpha}=\textrm{diag}
\left[1,\;\exp(-\frac{2i\pi\alpha}{M}),\;\exp(-\frac{4i\pi\alpha}{M}),\;\dots,
\:\exp(-\frac{2i(N-1)\pi\alpha}{M})\right].
\end{equation}

The randomness of the transform comes from the matrices $V$ and
$V^t$, where $[\dots]^t$ indicates the transpose matrix of the
matrix $[\dots]$. The matrix $V$ is generated by $N$ orthogonal
vectors $\{v_1,v_2,...,v_N\}$ as
\begin{equation}
V=[v_1,\; v_2,\; \dots, \; v_N].
\end{equation}
Where $v_n(n=1,2,...N)$ are column vectors which are the
eigenvectors of a symmetric random matrix $Q$ with
$Q_{mn}=Q_{nm}$. The matrix $Q$ can be obtained by an $N \times N$
real random matrix $P$ with a relation as
\begin{equation}
Q=(P+P^t)/2.
\end{equation}
The matrix $V$ satisfy the relation of $VV^t=I$.

Even though the DFRNT results in a random vector, the transform
itself still has several excellent mathematical properties as
follows.

[1] Linearity.
$\mathcal{R}^{\alpha}(ax_1+bx_2)=a\mathcal{R}^{\alpha}x_1+b\mathcal{R}^{\alpha}x_2$.

[2] Unitarity. $\mathcal{R}^{-\alpha}=(\mathcal{R}^{\alpha})^*$.

[3] Index additivity.
$\mathcal{R}^{\alpha}\mathcal{R}^{\beta}=\mathcal{R}^{\beta}\mathcal{R}^{\alpha}=\mathcal{R}^{\alpha+\beta}$.

[4] Multiplicity. $\mathcal{R}^{\alpha+M}=\mathcal{R}^{\alpha}$.

[5] Parseval energy conservation theorem. $\sum\limits
_K|X_\alpha(K)|^2=\sum\limits _k |x(k)|^2$.

\subsection{DFRNCT, DFRNST and their properties}

The definitions of DFRNCT and DFRNST can be directly given from
the DFRNT. Similar to the definitions of DFrCT and DFrST, of which
the eigenvectors can obtained from the eigenvectors of the
DFrFT\cite{pei:2001}, the eigenvectors of DFRNCT and DFRNST are
the same with that of the DFRNT if the same symmetric random
matrix $Q$ are used. That means when we construct the DFRNCT and
DFRNST, we use the same method to generate the matrix
$V=[v_1,\:v_2,\:\dots,\:v_N]$. However, in the following
discussion, we denote the the eigenvectors of the DFRNCT and the
DFRNST by matrices
\begin{equation}
V_c=[c_1, \; c_2,\; \dots\;,c_N],\quad V_s=[s_1, \; s_2, \;\dots,
\;s_N].
\end{equation}
Bear in mind that $c_n=s_n=v_n$ for the same matrix $Q$.

Similarly the kernel matrices of the DFRNCT and the DFRNST are
defined as follows
\begin{equation}
R_c^{\alpha}=V_cD_c^{\alpha}(V_c)^t, \quad
R_s^{\alpha}=V_sD_s^{\alpha}(V_s)^t.
\end{equation}
However the two eigenvalue diagonal matrices for DFRNCT and
DFRNST, {\it i.e.} $D_c^{\alpha}$ and $D_s^{\alpha}$ are,
respectively, chosen as follows
\begin{equation}
D_c^{\alpha}=\textrm{diag}\left[1,
\;\exp(-\frac{4i\pi\alpha}{M}),\; \dots,
\;\exp(-\frac{2(2N-2)i\pi\alpha}{M})\right],
\end{equation}
\begin{equation}
D_s^{\alpha}=\textrm{diag}\left[\exp(-\frac{2i\pi\alpha}{M}),\;
\exp(-\frac{6i\pi\alpha}{M}),\;\dots,
\;\exp(-\frac{2(2N-1)i\pi\alpha}{M})\right].
\end{equation}

The eigenvalue matrices of the DFRNCT and DFRNST are also diagonal
matrices with the diagonal elements chosen within the set of
$\{\exp(-2i\pi n\alpha/M):n=0,1,2,...2N-1\}$. The elements of the
DFRNCT are formed by the values with $n=2j,j=0,1,2,\dots$ while
the elements of the DFRNST are given by the values with $n=2j+1$.
The parameter $\alpha$ indicates the fractional order of the
transform and $M$ relate with the periodicity.

The DFRNCT and DFRNST of a 1-D signal $x$ and a 2-D image $y$ then
can be expressed as
\begin{equation}
X_c^{\alpha}=R_c^{\alpha}x, \quad X_s^{\alpha}=R_s^{\alpha}x,
\end{equation}
\begin{equation}
Y_c^{\alpha}=R_c^{\alpha}y[R_c^{\alpha}]^t, \quad
Y_s^{\alpha}=R_s^{\alpha}y[R_s^{\alpha}]^t.
\end{equation}
When $\alpha=0$, the kernels of the DFRNCT and DFRNST are identity
matrices. The DFRNCT is an identity matrix for $\alpha=M/2$,
however the DFRNST is a negative identity matrix ($R_s^{M/2}=-I$)
for $\alpha=M/2$.

From the Eq. 9 and Eq. 10 one can see that the DFRNCT and DFRNST
defined here are not generated by merely taking the real and
imaginary parts of the DFRNT kernel matrix. Therefore, the
construction of DFRNCT and DFRNST are different from the FrCT and
FrST in the Ref. \cite{Lohmann:1996}. If we define the DFRNCT and
DFRNST that way, they would lose the property of additivity, and
they would be a discrete transform in real number domain. And more
significantly in such cases the corresponding inverse transforms
do not exist.

On the definitions of DFRNCT and DFRNST, we adopt the method of
Pei and Yeh\cite{pei:2001}. Because we use orthogonal eigenvectors
to define the DFRNCT and DFRNST, the mathematical properties of
DFRNCT and DFRNST are thus similar to DFRNT.
\begin{itemize}

\item \textbf{Unitarity}. This property is directly from the
diagonal matrix $D_c^{\alpha}$ (and $D_s^{\alpha}$), that is
\begin{equation}
[R_c^{\alpha}]^*=[R_c^{\alpha}]^{-1}=R_c^{-\alpha},
\end{equation}
\begin{equation}
[R_s^{\alpha}]^*=[R_s^{\alpha}]^{-1}=R_s^{-\alpha}.
\end{equation}

\item \textbf{Index additivity}. This property always valid for
DFRNCT and DFRNST because
\begin{equation}
R_c^{\alpha}R_c^{\beta}=R_c^{\beta}R_c^{\alpha}=R_c^{\alpha+\beta}.
\end{equation}
\begin{equation}
R_s^{\alpha}R_s^{\beta}=R_s^{\beta}R_s^{\alpha}=R_s^{\alpha+\beta}.
\end{equation}
Moreover, according to this property, the matrices of inverse
transform of DFRNCT and DFRNST with fractional order $\alpha$ are
$R_c^{-\alpha}$ and $R_s^{-\alpha}$, respectively.

\item \textbf{Periodicity}. The DFRNST is periodic with $M$,
however the DFRNCT is periodic with $M/2$, {\it i.e.}
\begin{equation}
R_c^{\alpha+M/2}=R_c^{\alpha},\quad R_s^{\alpha+M}=R_s^{\alpha}.
\end{equation}

\item \textbf{Energy conservation theorem}.
\begin{equation}
\sum_q |R_c^{\alpha}x|^2=\sum_m |R_s^{\alpha}x|^2=\sum_n |x(n)|^2.
\end{equation}

\end{itemize}

\subsection{The relationships between DFRNCT, DFRNST and DFRNT}

The general relationships between DFRNT, DFRNCT and DFRNST can be
explored from the relationships of their eigenvalue matrices,
which are given by the equations Eq.~4, Eq.~9 and Eq.~10. All the
transforms take the same eigenvector matrices, the only
differences are their eigenvalues. From the definitions, one can
find immediately that
\begin{equation}
R_c^{\alpha}=\mathcal{R}^{2\alpha},\quad R_s^{\alpha}=\exp(-2i
\alpha \pi/M)\mathcal{R}^{2\alpha}.
\end{equation}
Therefore, the DFRNCT and DFRNST can be regarded as DFRNT's with
changes of scale in the fractional orders. In this instance, the
DFRNCT and DFRNST form two subsets within the whole domain of
DFRNT.

In above discussion we assume that all the transform kernels have
the same $N \times N$ dimension. We can also regenerate a DFRNT
with all the eigenvectors and eigenvalues of DFRNCT and DFRNST.
Such reconstructed DFRNT will have a $2N \times 2N$ or
$(2N+1)\times (2N+1)$ transform kernel matrices which can process
the data with $2N$- and $(2N+1)$-points long. In order to indicate
the difference between the reconstructed DFRNT and the original
definition, we refer to it as ReDFRNT. We assign the following
eigenvalue matrices to a ReDFRNT
\begin{equation}
D_{2N}^{\alpha}={\rm
diag}\left[1,\;\exp(-\frac{2i\pi\alpha}{M}),\;\dots, \;
\exp(-\frac{2(2N-1)i\pi\alpha}{M})\right],
\end{equation}
and
\begin{equation}
D_{2N+1}^{\alpha}={\rm
diag}\left[1,\;\exp(-\frac{2i\pi\alpha}{M}),\;\dots,\;
\exp(-\frac{4Ni\pi\alpha}{M})\right].
\end{equation}

For the $2N$-point ReDFRNT, the eigenvectors is given by the
eigenvectors of DFRNCT and DFRNST as the following matrix
\begin{equation}
V_{2N}=\frac{1}{\sqrt{2}}\left[\begin{array}{*{7}{c}} c_1,&
s_1, & c_2, & s_2, & \dots, & c_N, & s_N \\
c_1^z, & -s_1^z, & c_2^z, & -s_2^z,& \dots, & c_N^z, & -s_N^z\\
\end{array}\right].
\end{equation}
Where $c_n^z$ (or $s_n^z, n=1,2,...,N$) denotes the flipping
eigenvector $c_n$ (or $s_n$) in up-down direction. The
eigenvectors $V_{2N}$ are orthogonally vectors, {\it i.e.}
$V_{2N}V_{2N}^t=I$. For a $(2N+1)$-point ReDFRNT, however, the
eigenvectors should be expressed as
\begin{equation}
V_{2N+1}=\frac{1}{\sqrt{2}}\left[\begin{array}{*{9}{c}}
c_1, & s_1, & c_2, & s_2, & \dots, & c_N, & s_N, & v_0&\\
0, & 0, & 0, & 0, & \dots, & 0, & 0, & \sqrt{2}\\
 c_1^z, & -s_1^z, & c_2^z, & -s_2^z, & \dots, & c_N^z, &- s_N^z, & v_0 \\
\end{array}\right],
\end{equation}
so that $V_{2N+1}$ can also be an orthogonally eigenvectors, where
$v_0$ is zero vector. Then the kernel matrix of the ReDFRNT can be
written as
\begin{equation}
R^{\alpha}=V_mD_m^{\alpha}V_m^t,
\end{equation}
where $m$ denotes $2N$ or $2N+1$, respectively.

Obviously the ReDFRNT is different from the DFRNT originally
defined. Here the ReDFRNT, similar to the DFRNCT and DFRNST, can
be regarded as a subset of DFRNT. Such relationships resemble the
self-affine characteristics of fractals.

The eigenvector matrices of the ReDFRNT can also be written as the
format of column vectors, for instance $V_{2N}=\left[v_1 \; v_2 \;
\dots \;v_{2N}\right]$, where
\begin{equation}
v_{2n-1}=\left[
\begin{array}{c}
c_n \\
c_n^z\\
\end{array}\right],\quad
v_{2n}=\left[
\begin{array}{c}
s_n \\
-s_n^z\\
\end{array}\right],\;\:n=1,2,\dots,N.
\end{equation}

The eigenvectors $v_{2n-1}$ and $v_{2n}$ are even and odd
eigenvectors which have even and odd symmetry, respectively. Same
results are also valid for the matrix $V_{2N+1}$. Because the
eigenvectors of the ReDFRNT have well symmetry, the outputs can be
calculated with DFRNCT and DFRNST for even and odd signals,
respectively. The amplitudes of the ReDFRNT have even symmetry for
both of even and odd signals, and its phases have even symmetry
for the even signal. However, for the odd signal the symmetric
relationship of phases is not obvious because
\begin{equation}
\phi(n)=\phi(2N+1-n)\pm \pi.
\end{equation}

Where $\phi(n)$ is the phase of the ReDFRNT, $n$ can be taken the
value from $1$ to $2N$. We then define a special phase for the
ReDFRNT of an odd signal as

\begin{equation}
\phi'(n)=\arctan\left\{\tan[\phi(n)]\right\}, \
\phi'(n)\in[-\pi/2,\pi/2],
\end{equation}

Such defined special phase $\phi'(n)$ will be symmetric for an odd
signal, as depicted in Fig.~3.

The above proposition can be proved from the symmetry of the
eigenvectors of the ReDFRNT. Numerical calculations can also
demonstrate this proposition (see Fig.~1 to Fig.~3). The
symmetrical properties will be useful in the practical computation
of the reconstructed DFRNT. The eigenvectors of the DFRNCT and
DFRNST can be chosen as uniformly distributed values, thus the
computational load can be reduced to one half.

We know that in the continuous case, a function $f(x)$ can be
decomposed into an even and an odd functions($f_e(x)$ and
$f_o(x)$) as
\begin{equation}
f_e(x)=\frac{[f(x)+f(-x)]}{2},\ \ f_o(x)=\frac{[f(x)-f(-x)]}{2}.
\end{equation}
Similarly a discrete $2N$-point signal $s(n)$ can also be
decomposed into an even and an odd signals($s_e$ and $s_o$) as
\begin{equation}
s_e(n)=\frac{[s(n)+s(2N+1-n)]}{2},\ \ n=1,2,...,2N,
\end{equation}
\begin{equation}
s_o(n)=\frac{[s(n)-s(2N+1-n)]}{2},\ \ n=1,2,...,2N.
\end{equation}

Because the amplitude and the phase of a ReDFRNT have well
symmetry for even and odd signals, we only need to compute the
DFRNCT and the DFRNST of half of signal($s_e'(n)=s_e(n)$ and
$s_o'(n)=s_o(n)$, $n=1,2,...,N$) as
\begin{equation}
S_{ec}=R_c^{\alpha}s_e',\ \ S_{os}=R_s^{\alpha}s_o'.
\end{equation}

And the output $S(n)$ of the ReDFRNT can be computed with $S_{ec}$
and $S_{os}$ as

\begin{equation}
S(n)=\left\{%
\begin{array}{ll}
    S_{ec}(n)+S_{os}(n) \quad & \textrm{if $1 \leq n \leq N$,} \\
    S_{ec}(2N+1-n)-S_{os}(2N+1-n) \quad & \textrm{if $N < n \leq 2N$,} \\
\end{array}%
\right.
\end{equation}

For $(2N+1)$-point signal $s(n)$, the output of ReDFRNT can been
expressed this equation as follows
\begin{equation}
S(n)=\left\{%
\begin{array}{ll}
S_{ec}(n)+S_{os}(n) \quad & \textrm{if $1 \leq n \leq N$,}\\
s(N+1)\exp(-\frac{4Ni\pi\alpha}{M}) \quad & \textrm{if $n=N+1$,}\\
S_{ec}(2N+2-n)-S_{os}(2N+2-n)  \quad & \textrm{if $N+1 < n \leq
2N+1$.}\\
\end{array}
\right.
\end{equation}

Thereby, we can use an $N\times N$ kernel matrix to count the
ReDFRNT of an arbitrary $2N$-point(or $(2N+1$)-point) discrete
signal.

\section{Numerical results of simulation}

We know that the ReDFRNT, restructured by the DFRNCT and DFRNST,
provides symmetric distribution for both even and odd signals.
Such property can be easily demonstrated by the following
numerical calculations.

In the computation we choose the following two simple functions
$x_1(n)$ and $x_2(n)$ with size of $128$ points,

\begin{equation}
x_1(n)=\left\{%
\begin{array}{ll}
    1  \quad & \textrm{if $49 \leq n \leq 80$,} \\
    0  \quad & \textrm{otherwise,} \\
\end{array}%
\right.
\end{equation}
and
\begin{equation}
x_2(n)=\left\{%
\begin{array}{cll}
    1  \quad & \textrm{if $49 \leq n \leq 64$,} \\
    -1  \quad & \textrm{if $65 \leq n \leq 80$,}\\
    0  \quad & \textrm{otherwise.} \\
\end{array}%
\right.
\end{equation}

Where $x_1(n)$ and $x_2(n)$ are even and odd symmetric signals,
respectively, with $1 \leq n \leq 128$. The random matrices $P_c$
and $P_s$ are generated with MATLAB, and the fractional order
$\alpha$ and $M$ are chosen as $\alpha=0.6$ and $M=1$,
respectively. In order to compare the numerical results of DFRNCT
and DFRNST with ReDFRNT, we only take one half of the above signal
$x_1$ and $x_2$ for DFRNCT and DFRNST, {\it i.e.} the signals
$x_1'=x_1(n)$ and $x_2'=x_2(n)$, $n=1,2,..,64$, respectively. The
corresponding eigenvalues are calculated according to Eq.~9 and
Eq.~10 with $N=64$. The matrices of the eigenvectors are computed
from the random matrix $Q$ (Eq.~6), indicated by Eq.~5 and Eq.~7.
The ReDFRNT's of the signals $x_1$ and $x_2$ are calculated from
the eigenvalues given by Eq.~20 and Eq.~21, with the corresponding
eigenvector matrices given by Eq.~22 and Eq.~23, respectively.

The results of one dimension DFRNCT and the DFRNST are given in
Fig.~1 and Fig.~2. The bold lines denote the amplitude of the
DFRNCT and DFRNST for the half signals $x_1'$ and $x_2'$, which
coincide well with the ReDFRNT's amplitudes of $x_1$ and $x_2$,
respectively, when $n \le 64$. The symmetric properties of ReDFRNT
for an even signal in both amplitudes and phases have been
revealed clearly in the figures. The phase of the ReDFRNT for
signal $x_2$ is not symmetric, however the special phase
$\phi'(n)$ defined by Eq.~27 can be totally symmetric for the
ReDFRNT of the odd signal $x_2$. Such result is shown in Fig.~3.
Where we can find both amplitude and phase $\phi'(n)$ are
symmetric with respect to the central line $x=64.5$.

For the case of two dimensional transforms, three binary images
with $128\times 128$ pixels $I_1$, $I_2$ and $I_3$, shown in
Fig.~4(a)-(c) respectively, are used. The image only contain
simple rectangular patterns which is equivalent to some
rectangular functions of $\textrm{rect}(x-x_0,y-y_0)$.
Fig.~4(d)-(f) illustrate the corresponding results of ReDFRNT.
Here only amplitudes are displayed. The same parameters are chosen
as the case of one dimensional transforms. The results show that
the ReDFRNT keeps the symmetry of input images.

\section{Conclusion}

In this paper, we proposed the discrete fractional random cosine
and sine transforms (DFRNCT and DFRNST) based on the discrete
fractional random transform (DFRNT). These two random transforms
are generated by rearranging the distributions of eigenvalue
matrices meanwhile keeping the eigenvector matrix unchanged. Such
defined DFRNCT and DFRNST have excellent mathematical properties
inherited from DFRNT. We have demonstrated that the DFRNCT and
DFRNST are nothing but scaled DFRNT and form two subsets of DFRNT.
From these two transforms we can also reconstruct another kind of
DFRNT (ReDFRNT) which combines DFRNCT and DFRNST together. The
ReDFRNT has a special feature that it has symmetric distributions
for even and odd signals in both amplitudes and phases.

The DFRNT, DFRNCT, DFRNST and ReDFRNT are discrete fractional
order transforms with intrinsic randomness. We have demonstrated
that the DFRNT can be applied in information security such as
image encryption and decryption. Further applications of these
random transforms in image processing, pattern recognition,
artificial intelligence and so on are left as open questions for
the community of information science.

\newpage

\section*{List of figure captions}

Figure 1. The results of DFRNCT and ReDFRNT of the 1-D signal
$x_1$, with $\alpha=0.6$ and $M=1$. Because the function $x_1$ is
even, so that the ReDFRNT are symmetrical and coincides with the
DFRNCT.

\vspace{1cm}\noindent Figure 2. The results of DFRNST and ReDFRNT
for the odd signal $x_2$ with $\alpha=0.6$ and $M=1$. The
amplitude of ReDFRNT are symmetric and coincides with the result
of DFRNST.

\vspace{1cm}\noindent Figure 3. The same calculation results of
DFRNST and ReDFRNT for the signal $x_2$ with special phase
$\phi'(n)$. Where we also take $\alpha=0.6$ and $M=1$. The special
phase then is symmetric distributed for the ReDFRNT.

\vspace{1cm}\noindent Figure 4. The results of ReDFRNT for 2-D
functions: (a), (b) and (c) display three input images $I_1$,
$I_2$ and $I_3$, respectively. From (d) to (f) are the
corresponding ReDFRNT's, with $\alpha=0.6$ and $M=1$, for images
$I_1$, $I_2$ and $I_3$, respectively. From the results we can find
that the ReDFRNT keep the symmetries in the input images.

\newpage

\begin{figure}[h]
\includegraphics[width=15.2cm]{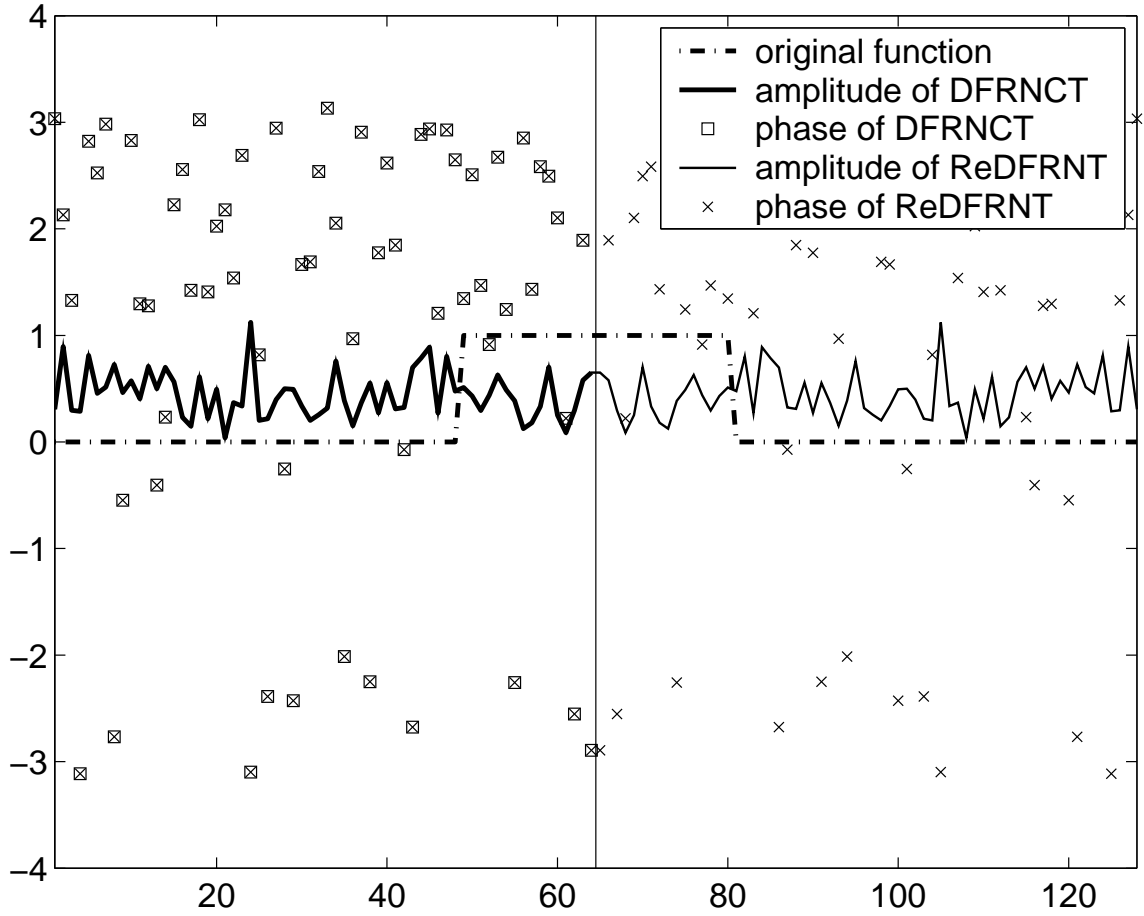}
\caption{The results of DFRNCT and ReDFRNT of the 1-D signal
$x_1$, with $\alpha=0.6$ and $M=1$. Because the function $x_1$ is
even, so that the ReDFRNT are symmetrical and coincides with the
DFRNCT.}
\end{figure}

\newpage

\begin{figure}[h]
\includegraphics[width=15.2cm]{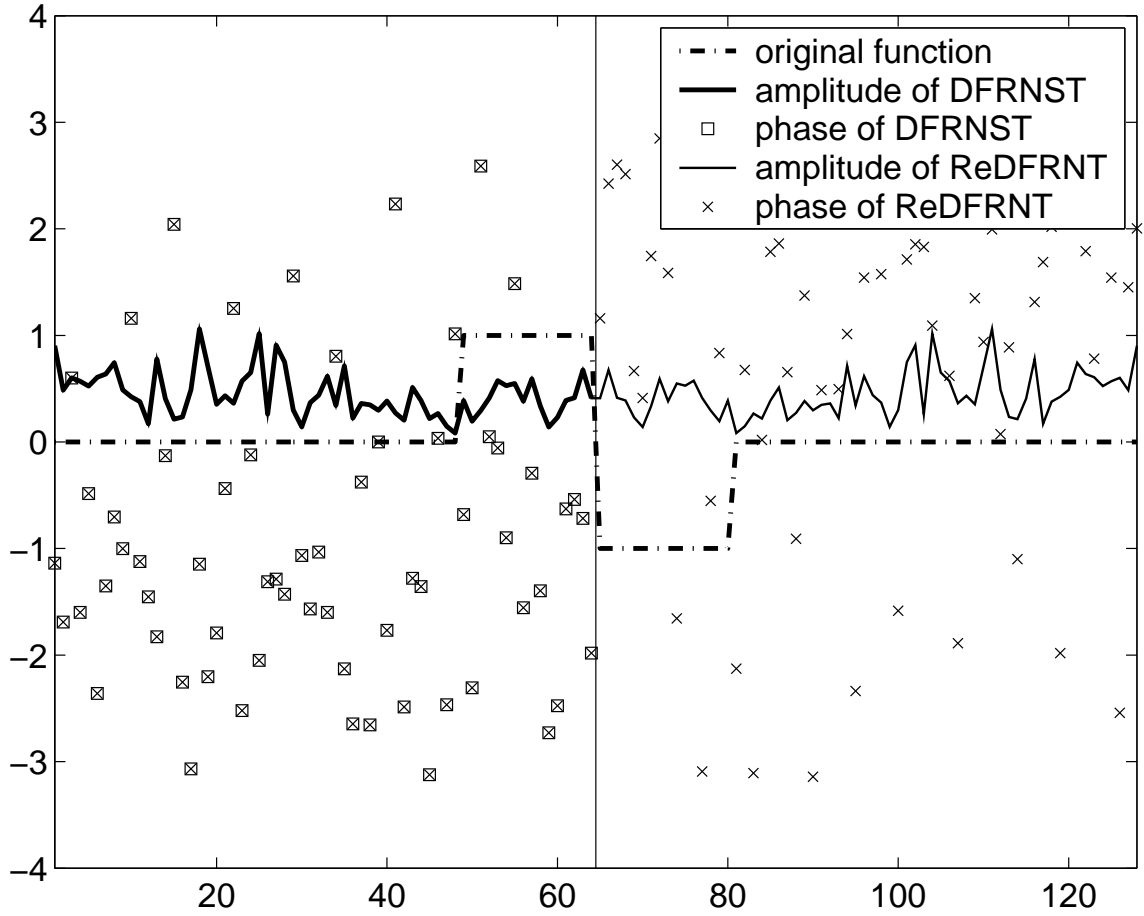}
\caption{The results of DFRNST and ReDFRNT for the odd signal
$x_2$ with $\alpha=0.6$ and $M=1$. The amplitude of ReDFRNT are
symmetric and coincides with the result of DFRNST.}
\end{figure}

\newpage

\begin{figure}[h]
\includegraphics[width=15.2cm]{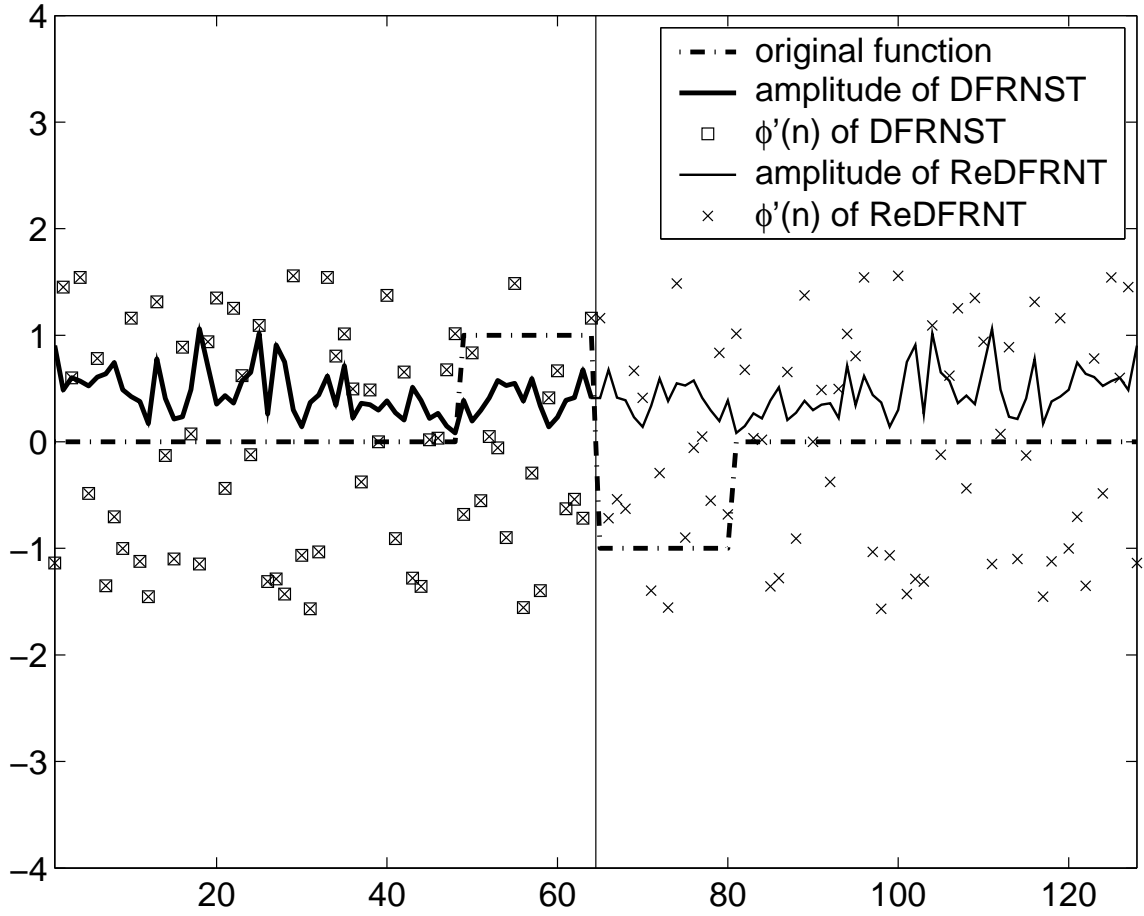}
\caption{The same calculation results of DFRNST and ReDFRNT for
the signal $x_2$ with special phase $\phi'(n)$. Where we also take
$\alpha=0.6$ and $M=1$. The special phase then is symmetric
distributed for the ReDFRNT.}
\end{figure}

\newpage

\begin{figure}[h]
\includegraphics[width=14cm]{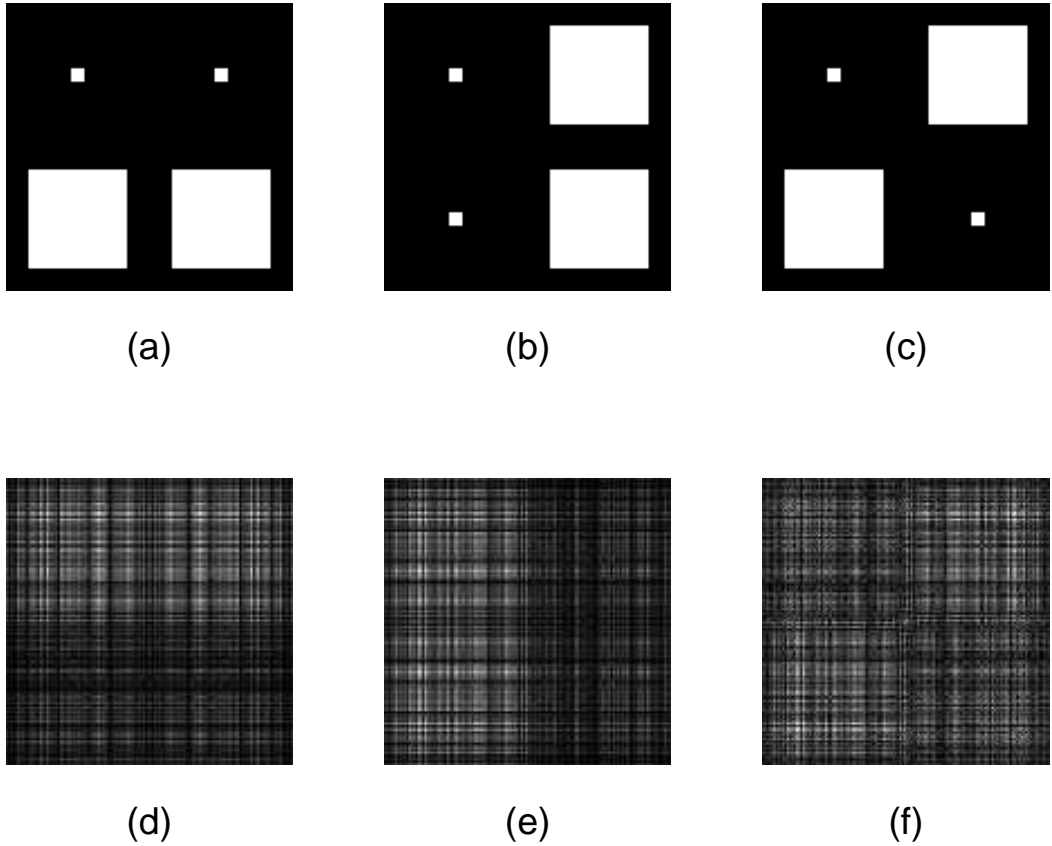}
\caption{The results of ReDFRNT for 2-D functions: (a), (b) and
(c) display three input images $I_1$, $I_2$ and $I_3$,
respectively. From (d) to (f) are the corresponding ReDFRNT's,
with $\alpha=0.6$ and $M=1$, for images $I_1$, $I_2$ and $I_3$,
respectively. From the results we can find that the ReDFRNT keep
the symmetries in the input images.}
\end{figure}

\end{document}